%% file: main.tex
\documentclass[journal]{IEEEtran}
\usepackage{amsmath}
\usepackage{amssymb}

\usepackage{graphicx}
\graphicspath{ {./figures/} }

\usepackage{caption}
\newcommand{\source}[1]{\caption*{Source: {#1}} }

\begin{document}
\title{Privacy Preservation Among Honest-but-Curious Edge Nodes: A Survey}

\author{Christian Badolato}

\maketitle

\input{abstract}
\input{introduction}
\input{privacy-vs-security}
\input{edge-computing}

\input{ec-and-privacy}
\input{conclusion}

\bibliographystyle{IEEEtran}
\bibliography{bibliography}

\end{document}

%% file: abstract.tex
% As a general rule, do not put math, special symbols or citations in the abstract or keywords.

\begin{abstract}
Users care greatly about preserving the privacy of their personal data gathered during their use of information systems. This extends to both the data they actively provide in exchange for services as well as the metadata passively generated in many aspects of their computing experiences. However, new technologies are at a great risk of being inadequate to protect a user's privacy if researchers focus primarily on the use cases of these technologies without giving sufficient consideration to incorporating privacy at a fundamental level. Edge computing has been introduced as a promising networking paradigm for processing the incredible magnitude of data produced by modern IoT networks. As edge computing is still considered a relatively new technology, the edge computing community has a responsibility to ensure privacy protection is interwoven into its implementations at a foundational level. In this paper, I first introduce the concepts of user privacy and edge computing; I then provide a state-of-the-art overview of current literature as it relates to privacy preservation in honest-but-curious edge computing. Finally, I provide future research recommendations in the hope that the edge computing research and development community will be inspired to ensure strong privacy protections in their current and future work. 
\end{abstract}


%% file: introduction.tex
\section{Introduction}

\IEEEPARstart{I}{nternet} of Things (IoT) devices have, quite literally, taken over the world in terms of the sheer number of connected devices and the volume of data they process. While the number of non-IoT devices has remained constant at roughly 10 billion connected devices since 2013, the quantity of IoT devices increased from 2.1 billion in 2013 to 10 billion in 2019 and is expected to grow to between 30.9 and 55.7 billion devices by 2025 \cite{statistaInternetThingsIoT2021} \cite{internationaldatacorporationFutureIndustryEcosystems2021}; these devices are expected to generate almost 80 zettabytes (or 80 trillion gigabytes) of data \cite{internationaldatacorporationFutureIndustryEcosystems2021}. Given this massive future growth, there are increasing concerns that this projected boom in IoT devices could overwhelm already existing cloud computing infrastructure due to strict requirements on storage, latency, and reliability, as well as other attributes \cite{abdulkareemIoTCloudComputing2021}. As a result, this device boom has necessitated the development of a new system architecture to handle the vastly increased quantity of data being processed and transferred; the developed architecture became known as edge computing. By grouping and connecting geographically local devices (both IoT and non-IoT) to a geographically local server-class machine, many cloud computing issues related to device volume, such as bandwidth constraints and latency considerations, can be alleviated.

The addition of intermediate processing nodes comes with newfound data privacy concerns. Information which was previously transmitted only to a trusted cloud server is now received by at least one additional machine which increases the potential to leak sensitive information. As the owners of the cloud server need not provision these intermediate machines, they may be considered untrusted by the devices which communicate with them. This presents a new challenge: how the privacy of a user's identity and data can be protected while leveraging the benefits of the edge computing paradigm despite transmission and processing involving untrusted nodes. Though researchers have proposed algorithms (such as the one in \cite{zhangPrivacyawareTaskAllocation2020}) which allow users to specify which information to share based on their perceived benefits, it has been repeatedly shown that users will act in a manner contradictory to their own stated privacy desires when they expect to receive an immediate benefit for sacrificing that privacy \cite{wilsonUnpackingPrivacyParadox2012}. Because of this, it is in users' best interest---and thus should be a main goal for research---that edge computing-based algorithms contain privacy protections as a fundamental aspect of their construction.

This paper provides an overview of the current state of privacy-preserving algorithms in edge computing. To my knowledge, this is the first edge computing survey focused solely on privacy rather than grouping security and privacy together, which tends to dilute the importance of privacy as its own consideration. The main contributions of this paper are the following:
\begin{itemize}
\item I define and differentiate security and privacy to promote an understanding on why researchers and algorithm designers must consider privacy independently of security.
\item I provide an overview of the edge computing paradigm and discuss its application domains and data processing paradigms.
\item I survey the current literature on privacy-preserving edge computing algorithms, provide insight where appropriate, and categorize the literature by several factors.
\item I state key takeaways from the surveyed literature and recommend paths for future research.
\end{itemize}

The remainder of this paper is organized as follows: Section \ref{privacy-vs-security} defines the concepts of privacy and security, contrasts the two, and provides a background for why privacy is particularly important. Section \ref{ec-overview} provides an overview of the edge computing paradigm and highlights current usages for the paradigm. Section \ref{survey} surveys current privacy literature involving honest-but-curious edge computing environments. Finally, Section \ref{conclusion} provides a summary and recommendations for future research directions.

%% file: privacy-vs-security.tex
\section{Privacy vs. Security} \label{privacy-vs-security}
Despite frequently being conflated in published literature, security and privacy are two distinct concepts in the field of Information Technology (IT). 

Information Security (IS) has many accepted definitions such as the United States Government's definition of ``protecting information and information systems from unauthorized access, use, disclosure, disruption, modification, or destruction in order to provide confidentiality, integrity and availability'' \cite{u.s.governmentpublishingofficeTitle44PUBLIC1968}, or the Information Systems Audit and Control Association's (ISACA) definition of ensuring ``information is protected against disclosure to unauthorized users (confidentiality), improper modification (integrity) and nonaccess when required (availability)'' \cite{ISACAInteractiveGlossary}. However, almost all widely accepted definitions of IS involve the protection of data from access, modification, or access denial stemming from unauthorized parties. This is the essence of IS---ensuring data is accessible only by those authorized by the data owner and that the data is valid and available for those authorized users. 

Data privacy focuses on the use of data by authorized parties; it defines how information is handled and among whom that information can be shared \cite{bambauerPrivacySecurity2013}. In general, privacy tends to be more policy focused and outlines which procedures information systems must follow in order to be compliant. Despite this, specific technical methods for privacy implementation do exist in the form technical privacy controls \cite{nistjointtaskforceSecurityPrivacyControls2020} and privacy-aware data processing algorithms. Overall, data privacy is about keeping the information which data owners do not want to disclose hidden while authorized parties and systems process and store that data. Put simply, the main difference between security and privacy in IT lie in  authorization, with security focusing on preventing access by unauthorized parties and privacy focusing on which data can be seen, stored, used, and transferred by authorized parties.

Data privacy is a significant concern for users of information systems. A survey of Americans conducted by PEW Research Center in 2019 showed that 79\% of surveyed adults were concerned about the use of their collected data by companies, 81\% believed the risks of data collection outweighed its possible benefits, and 81\% felt a lack of control over how their private data is currently protected \cite{pewresearchcenterAmericansPrivacyConcerned2019}. Therefore, it is crucial that researchers ensure that the protection of user privacy is pervasive through proposed designs, models, and algorithms going forward.

%% file: edge-computing.tex
\section{Edge Computing Overview} \label{ec-overview}
Edge computing is a network and system architecture that places servers (hereafter referred to as ``edge nodes'') between end user devices and the remote cloud servers (hereafter referred to as ``the cloud''). These edge nodes may be responsible for any combination of data processing, storage, or aggregation and are often used to offload a significant amount of data processing from the cloud \cite{shiEdgeComputingVision2016}. The paradigm is designed to minimize latency, increase privacy and security, delegate processing, and reduce the quantity of data transmitted to the cloud over long distances \cite{shiEdgeComputingVision2016} \cite{edgecomputingtaskIntroductionEdgeComputing2018}. The scalability of edge computing is also a significant advantage since many devices, limited only by the edge node's capacity, can be added to the network while adding only a single additional link---and therefore only a single link's worth of bandwidth requirements---to the cloud.

While the paradigm was first introduced by Akamai Technologies in the late 1990s with their implementation of content delivery networks (CDNs), edge computing as it is understood today started gaining popularity when Flavio Bonomi et al. introduced the concept of fog computing, an architecture similar to edge computing, in their paper \textit{Fog Computing and Its Role in the Internet of Things} \cite{bonomiFogComputingIts2012} \cite{satyanarayananEmergenceEdgeComputing2017a}. With the recent increase in internet-connected IoT devices, edge computing has become a particularly enticing paradigm for IoT networks; these devices often have heavily restricted computation power and on-board memory and are frequently designed purely to gather and offload data for external processing \cite{bonomiFogComputingIts2012}. Due to the incredible amount of data these IoT devices are expected to processes, it is critical to aggregate and analyze this data as close to these networks as possible to minimize bandwidth usage across the global networks. 

However, while IoT applications are a staple of edge computing, the model itself does not preclude non-IoT devices from taking advantage of the paradigm. Researchers are actively finding uses for edge computing across many domains involving several methods of data processing. In the remainder of this section, I highlight several application domains that use edge computing, followed by an overview of the main types of data processing algorithms and a definition of two terms which are becoming increasingly pervasive as a result of edge computing. 

\subsection{Application Domains}
Since the introduction of modern edge computing, researchers from many application domains have begun to explore the advantages of using edge computing. Here I highlight five domains with the most prolific uses of the paradigm.

\subsubsection{Medical}
Given the rapid increase in usage of IoT devices in the world of healthcare \cite{pradhanIoTBasedApplicationsHealthcare2021}, it is of little surprise that medicine is a primary field that explores the applications of edge computing. Examples of the ways in which the medical research community have embraced edge computing are through optimization of processing tasks involving real drug reaction data gathered from elderly patients' wearable sensors \cite{rajOptimizedMobileEdge2021}, COVID-19 diagnosis and symptom monitoring through machine learning of datasets gathered through sensors at a user's home \cite{rahmanInternetofMedicalThingsEnabledEdgeComputing2021}, and blockchain-based medical data exchange algorithms facilitated by edge computing \cite{awadabdellatifMEdgeChainLeveragingEdge2021}. 

\subsubsection{Vehicle Networks}
The Internet of Vehicles (IoV) and vehicular ad hoc networks (VANETs) are a large step forward in the pursuit of modern smart and autonomous vehicles providing lightweight and efficient computation and coordination services for highly connected vehicular infrastructure \cite{kimInternetVehiclesIoV2020} \cite{ghoriVehicularAdhocNetwork2018}. Within these technologies, edge computing helps facilitate operations such as data sharing \cite{cuiEdgeComputingVANETsAn2020}, task offloading optimization \cite{wangJointOptimizationTask2022}, and machine learning for autonomous operations \cite{liApplicationIOTDeep2021}.

\subsubsection{Industrial Control Systems}
Industrial automation may be one of the oldest domains of machine-based automation starting with the first programmable logic controller (PLC) created by General Motors in the late 1960s \cite{processsolutionsincorporatedBriefHistoryProgrammable2020}. Since then, manufacturing and industry have constantly sought innovations in control system autonomy to increase productivity, quality, efficiency, and safety \cite{frohmINDUSTRYVIEWAUTOMATION2006} \cite{surecontrolsincorporatedWhatIndustrialAutomation2013}. With it's ability to perform low-latency computations locally without the high bandwidth costs required to send terabytes of data to the cloud, edge computing has become a cornerstone technology in providing monitoring, analytics, and predictive maintenance recommendations to industrial control systems \cite{edgecomputingtaskIntroductionEdgeComputing2018}.

\subsubsection{Smart Cities and the Energy Internet}
As clean and efficient energy solutions become a focus for society, frameworks and network topologies for managing these energy networks become critical. Within these networks, solutions are needed for controlling various types of energy production, managing and tracking the consumption of energy, efficiently load-balancing energy distribution, and ensuring a smooth interface for users to request energy \cite{maOverviewEnergyInternet2016}. When it comes to edge computing, the natural breakdown of city and residential infrastructure allows edge computing to position itself as a natural paradigm for this field; countries are often broken into subdivisions such as provinces, states, or prefectures which can further be broken into elements such as counties, municipalities, boroughs, and eventually individual buildings. This hierarchical structure allows edge nodes to handle varying levels of processing at each tier of infrastructure. 

Within this framework, researchers are already examining edge computing based methods for tracking energy consumption, providing bidding and trading services for energy surplus, and ensuring the security of energy usage data \cite{guanECOSECURITYTacklingChallenges2019} \cite{liNovelEdgeComputing2021}. 

\subsubsection{Location-based Services}
The utility and convenience of location-based services has been pervasive in society since the inception of GPS-supported mobile devices \cite{bellavistaLocationBasedServicesBack2008}. The high use of navigation applications such as Google Maps and Apple Maps, trip planners such as Trip Advisor and Yelp, location-based social media features such as Facebook's Check-In system and Foursquare, and many other types of location-based services show that many users greatly value the convenience provided by customized directions and recommendations based on their current and historical locations \cite{andersonMoreAmericansUsing2016}. Edge computing provides the ability to operate on this data locally which can help preserve location privacy by negating the need to upload this data to the cloud \cite{shiEdgeComputingVision2016}, enhance computational efficiency by pre-allocating the required offloading resource requirements based on predicted locations \cite{romanMobileEdgeComputing2018}, or provide more accurate recommendations based on the profiles of users within a given location context \cite{zhouDifferentiallyPrivateTrustworthyOnline2019}.

\subsection{Data Processing Paradigms}
As evidenced by the large variation in application domains using edge computing, there are several underlying end goals in edge computing processing. Here I introduce four common data processing paradigms---which I will survey in Section \ref{survey}---as it relates to the edge computing paradigm. 

\subsubsection{Data Transmission}
Perhaps the most straightforward application of any multi-node data processing algorithm is that of data transmission. These algorithms are focused solely on sending data from one end user to a server or other end user in such a way that the original data can be easily retrieved by an interested, authorized party. For the purpose of this paper, this retrieval need not be immediate and data storage algorithms will be included under this umbrella. Though not required to be considered a data transmission algorithm, most of the surveyed methods of data transmission include protection to ensure the data preserves its integrity during transmission and that the original data can only be received or recovered by authorized parties.

\subsubsection{Data Aggregation}
The purpose of data aggregation is almost identical to that of data transmission. In aggregation, however, the original data need not (and often should not) be able to be retrieved or recovered by the destination device. Instead, aggregation focuses on taking data from numerous sending devices and combining it into a summarization derived from the originals \cite{ibmIBMDocs2021}. Data aggregation is often used to gather meaningful metrics that can be analyzed and acted upon while preserving the privacy of individual data producers; for example, in the medical field, it can be valuable to know what percentage of Hispanic males aged 50--59 have been diagnosed with skin cancer without being able to infer that a particular user X is a Hispanic male aged 50--59 with skin cancer.

\subsubsection{Federated Learning}
Federated Learning is a relatively new machine learning technique---first established by Google in 2016---that enables training a single machine learning model with data residing on any number of distributed end devices. These devices may either produce their own data sets or be given a subset of a global data set on which to train. This approach can help improve data privacy since end devices need not share their data sets with other devices or even a central server, and can provide more efficient training as the model training process is no longer limited to the processing power of a single device \cite{mcmahanCommunicationEfficientLearningDeep2017}.

Though, while federated learning can improve data privacy, the process itself does not guarantee it to an adequate degree. While keeping a dataset locally without needing to share it across the network has been shown to significantly increase the degree of privacy in practice \cite{liuKeepYourData2021}, attacks against training model parameters which can reconstruct a useful, if not accurate, representation of the original dataset exist, which can cause privacy leakage even in federated models \cite{geipingInvertingGradientsHow2020}.

\subsubsection{Process Offloading}
Processing and decision offloading allows devices with insufficient resources to delegate their processing tasks to a device which possesses the required resources and knowledge to perform the tasks with minimal latency. Unlike traditional methods of processing delegation which use well defined rules to delegate tasks to servers within the device's immediate computing group, offloading allows tasks to be distributed among arbitrary nodes, networks, and distances provided the receiving node is both able to handle the request and known to the originating device \cite{kumarSurveyComputationOffloading2013}. Beyond resource considerations, decision offloading may also be performed to allow devices with incomplete knowledge to request a decision from a device with complete knowledge. This is particularly relevant in machine learning environments in which a single, central node possesses a trained model capable of being queried from resource-light end devices \cite{maLightweightPrivacypreservingMedical2020}.

Furthermore, offloading is not reserved strictly for end devices offloading tasks to a central server. Tasks may also be offloaded from a central server (whose role is to collect information on the tasks which must be completed) to end devices to perform distributed computations, execute simultaneous processing, and collect multiple data points which the central server may not have been able to collect itself \cite{dingPrivacypreservingTaskAllocation2022}. Regardless of the purpose or framework behind the method of task allocations, offloading provides collaborative decision-making and processing ability beyond the limitations of a single device.

\subsection{Crowdsensing and Crowdsourcing}
Crowdsensing and crowdsourcing (also referred to as mobile crowdsensing and mobile crowdsourcing), are two similar, but different, data acquisition paradigms through which the value of edge computing shines. In crowdsensing, users opt to gather and produce data based on their own locally sensed environments which is then used for either machine learning model training (federated or centralized), data aggregation, or pure transmission. Crowdsourcing, meanwhile, is a specialized form of offloading in which a central server or edge node requests for distributed group of end devices to perform tasks that would either be too computationally expensive for the server to process on its own or which requires knowledge that the server does not possess.

Though not unique to edge computing, the growing popularity of the paradigm has been hastening the implementation of crowdsensing and crowdsourcing in the modern world \cite{yuCrowdsensing2021} \cite{angCrowdsourcingInternetThings2022}. Crowdsensing and crowdsourcing provide their own unique privacy challenges to the world of edge computing as end user devices may be capable of collecting highly sensitive data \cite{wangWhenMobileCrowdsensing2019} \cite{xiaPrivacyCrowdsourcingReview2020}. Therefore, special care must be taken when looking at algorithm designs to ensure user privacy is preserved. I will highlight papers focusing on crowdsensing applications in Section \ref{survey}. No surveyed papers relate directly to crowdsourcing, therefore, crowdsourcing will not be discussed.

%% file: ec-and-privacy.tex
\section{Privacy Preservation among Honest-but-Curious Edge Nodes} \label{survey}
Even among similar edge computing architectures, the expected threat model for a system can vary wildly. Assumptions that edge nodes are owned by a single, trusted entity, multiple trusted entities, malicious entities, unknown entities, or any other form of ownership structure are all present in modern literature. As a result, it is imperative to establish a threat model before considering the efficacy of privacy-conscious edge computing algorithms. 

In this survey, I focus on the honest-but-curious edge node model as defined by Paverd et. al \cite{paverdModellingAutomaticallyAnalysing2014b}; within this model, all edge nodes behave honestly---they will not deviate from the established communication and storage protocols they are expected to follow---but, they will attempt to learn all they can about end devices' data and may collude among each other, central servers, or other end devices to relate sensitive information to the specific end device to which it belongs. In this model, edge nodes are not considered trusted. While they may be freely used to perform data processing and task allocations without concern for the validity of the operations, a device within the system must never provide an edge node with the means to determine a link between decipherable data and the end device which produced it. Here, the, distinction of "decipherable" data is key --- edge nodes may be allowed to associate encrypted data with a particular data producer as long as it does not gain the means to provide itself or any other device access to the decrypted data. I do not concern myself with the trustworthiness of other nodes within the context of this paper, however, many surveyed algorithms include protections against semi-trusted or untrustworthy cloud servers or end devices.

In addition to the chosen threat model, preservation of privacy in edge computing is highly dependent on the final purpose of the data flowing through the system. In this section, I survey current literature to analyze state-of-the-art algorithms for preserving user privacy within an honest-but-curious edge node framework with respect to the data processing paradigms defined in Section \ref{ec-overview}. These algorithms are classified into several overarching categories which are highly prevalent in recent literature; these categories are not mutually exclusive and many proposed algorithms incorporate aspects from two or more categories as is summarized in Table \ref{table:techniques:1} and Table \ref{table:techniques:2}. Table \ref{table:data-uses:1} and Table \ref{table:data-uses:2} summarizes the target data processing paradigm for each algorithm. Finally, Table \ref{table:crowdsense-source} provides a quick reference for literature involving crowdsensing since, given the growing popularity of this concept, the risk of privacy leakage within it warrants special attention.

\input{tables/techniques}

\subsection{Trusted Third Party}
A common element adopted by many security architectures in data transmission and processing is a Trusted Third Party (TTP), and edge computing is no exception. Often used for generating and distributing security parameters such as public and private keys, shared secrets, and transaction verification information, TTPs can provide an avenue for performing cryptographic operations in frameworks where end devices do not contain the required processing or storage capabilities to perform these operations themselves \cite{Adams2011}. TTPs provide a centralized authority for validating and verifying transactions and are required to be trusted by all parties who will take part in communication and data storage. Especially in the realm of resource-constrained IoT devices, as seen below, many privacy-preserving edge computing algorithms rely on TTPs for verification and validation of data and transmissions to provide strong security guarantees while still preserving user privacy. As the honest-but-curious model applies only to edge nodes in the context of this paper, it can be assumed that a TTP can be fully trusted.

As many algorithms which fall in other categories utilize TTPs in some part of their design, the papers which appear in this section utilize TTPs as their primary method of ensuring data privacy. Papers surveyed under other techniques may contain one or more TTPs in their implementation, however, they are not the primary focus of their respective algorithms. Refer to Table \ref{table:techniques:1} and \ref{table:techniques:2} for a full summary of applicable techniques present within each paper.

The authors of \cite{zengMMDAMultidimensionalMultidirectional2020} define an enhanced data aggregation scheme using a TTP for key generation. They note that, while most multi-dimensional implementations let the cloud compute the aggregation of multiple device readings for the same type of reading (for example, the summation of the first vector index for all devices), it is often valuable to be able to aggregate all readings for a single device as well (for example, compute the summation of all elements in a device’s data vector). The authors seek to address the deficiency in supported aggregation calculations in current multi-dimensional data aggregation implementations. To allow this calculation, they include a TTP which generates both individual data atom keys for the end devices and aggregate column and row keys for the cloud server. By this method, each individual data atom read by an end device can be altered by a unique blinding factor \cite{bleumerBlindingTechniques2005}, transmitted to the edge nodes, and aggregated into both rows (one per device) and columns (one per reading type). The aggregates can then be transmitted to the cloud server which can then unblind the row and column aggregates.

Hsu et. al \cite{hsuPrivacypreservingDataSharing2020} also use a TTP to generate keys from which an IoT user or device can compute their own public and private keys for data sharing between IoT users and IoT devices. These keys are constructed to allow the IoT devices to compute a shared symmetric encryption token generated by the IoT users for data transmission without revealing this token to an edge node. As it provides the foundation of TLS communication, the use of public and private keys to encrypt a shared symmetric key is already a common concept in security and privacy known as hybrid encryption \cite{networkworkinggroupTransportLayerSecurity2008}, however, the uniqueness of Hsu et. al's approach is in computing feature vectors for data search queries from the IoT users in such a way that they can be used by attribute-based encryption schemes \cite{chaudhariReviewAttributeBased2016} which prevent the decryption of the query unless the attributes of the feature vector adequately match the provided attributes of an IoT device. This allows a search algorithm to be performed by the edge node to determine which IoT devices should receive the encrypted symmetric key. The IoT devices can then perform secure communication with the users through an edge node using the shared symmetric key.

The work produced by Hsu et. al \cite{hsuReconfigurableSecurityEdgeComputingBased2018} is a unique proposal worth highlighting in the context of this paper. Though their paper itself focuses on security and not necessarily data privacy, the authors provide a mechanism for providing fully anonymous authentication and Attribute-based Access Control (ABAC) among edge-based IoT networks through the addition of a global key management server (acting as the TTP) and an anonymous authentication, authorization, accounting (AAA) framework. As identification is required to link private data with an end device, the ability to provide authentication and ABAC without requiring consistent identifiers is a major step towards complete data privacy.

While the above papers demonstrate the power of TTPs as cryptographic tools in resource-constrained environments, the criticisms of TTPs abound. Being centralized, these elements can still require significant bandwidth allocation to adequately process requests with minimal latency if poorly implemented, which has the potential to counteract many benefits of edge computing overall. Additionally, TTPs require experience, precision, and care to setup and maintain which can cause both privacy and security holes if implemented poorly and can be seen as single points of failure for privacy protection since collusion between a TTP and an unauthorized third party could cause severe privacy leakage within the system \cite{Adams2011} \cite{leviProblemTrustedThird2002}.

\input{tables/techniques-two}

\subsection{Homomorphic Encryption}
Homomorphic encryption allows for mathematical operations to be performed on encrypted data in such a way that these operations are reflected in the unencryption of the data. Though encryption algorithms which support arbitrary operations on the data (called fully homomorphic encryption, or FHE) is still a topic of heavy research due to unreasonable  resource and time requirements \cite{marcollaSurveyFullyHomomorphic2022}, partially and somewhat homomorphic schemes, which both support the manipulation of ciphertext through a limited subset of operations (typically addition and multiplication, called additive and multiplicative homomorphism respectively) and across a limited number of operations (for example, the Boneh-Goh-Nissim system \cite{bonehEvaluating2DNFFormulas2005} can only support a single multiplicative operation), are actively used in many privacy-preserving edge computing algorithms.

Private data aggregation is a key use of homomorphic encryption in edge computing and, as it often does not require any additional computation or coordination beyond the encryption, aggregation, and decryption of the homomorphic data itself, has been shown to be a lightweight solution. The authors of \cite{liPrivacyPreservingData2019} use the Boneh-Goh-Nissim system to allow end devices to encrypt data and transmit it to an edge node in a manner that allows for integrity checking without revealing any plaintext to the edge nodes or allowing any identity-to-data linkage for the cloud server. Ma et. al \cite{maEdgeComputingAssisted2021} and Zhang et. al \cite{zhangLPDAECLightweightPrivacyPreserving2018} both explore similar methods of aggregation utilizing the Paillier homomorphic cryptosystem \cite{paillierPublicKeyCryptosystemsBased1999} with their data integrity verification algorithms being their primary difference.

Li et. al \cite{liLightweightFineGrainedPrivacyPreserving2021} propose an expanded, Paillier-based scheme allowing the cloud server to supply fine-grained data aggregation rules that dictate which nodes' data should be included in which aggregates. It is noted that in this scheme there is no protection in place to ensure that these rules do not contain only a single end device and, therefore, the cloud server can isolate a single end device's data. However, as the edge node remains incapable of viewing the data in plaintext, the honest-but-curious edge node model is still satisfied. All the above schemes noted performance improvements over other privacy-preserving aggregation schemes.

Homomorphic encryption is also widely used in machine and federated learning. The authors of \cite{liVerifiablePrivacypreservingMachine2021} detail a method for performing linear regression model predictions based on an end-device supplied data vector encrypted by the OU encryption scheme \cite{okamotoNewPublickeyCryptosystem1998}. Through this method, they can perform the prediction while keeping the model private to the cloud server, the data private to the end devices, and without leaking any privacy data to the edge nodes. Li et. al \cite{liPrivacyPreservedFederatedLearning2021} focuses on federated model training among connected vehicles by having the cloud server transmit a learning model to the end vehicles and then allowing edge servers to aggregate Dijk-Gentry-Halevi-Vaikutanathan \cite{vandijkFullyHomomorphicEncryption2010} encrypted model parameter updates sent from the vehicles weighted by participant reputation. These encrypted aggregates are periodically sent to the cloud server to update the overall model.

In the realm of task allocation, the authors of \cite{dingPrivacypreservingTaskAllocation2022} detail a method for performing private crowdsensing task allocation through comparing Paillier encrypted distances without revealing actual location data as the distance calculation can both perform and deconflict task allocations on completely encrypted location data. Additionally, by adding a degree of random noise to the encrypted data, the actual location data of an end device can be hidden from the cloud server. The authors of \cite{maPrivacyPreservingReputationManagement2019} handle crowdsensing reputation management by proposing an algorithm for computing the ranking of the standard deviations of sensing data encrypted by Fan and Vercauteren's somewhat-homomorphic encryption algorithm \cite{fanSomewhatPracticalFully2012}; the homomorphic properties of the encryption allow for the computation of an encrypted standard deviation which can then be compared and ranked to update an end device's reputation without revealing its individual sensing data.

\input{tables/data-uses}

\subsection{Differential Privacy}
First introduced by Dwork et. al in 2006, $\epsilon$-differential privacy provides a constraint on the data a randomized algorithm can return to limit the amount of potentially private information a dataset query can provide. An $(\epsilon, \delta)$-differentially private algorithm is one where the result of the algorithm's execution of a given query on any applicable dataset is nearly indistinguishable from the result of the same query when performed on a dataset that differs from the original by the addition or removal of a single entry \cite{dworkCalibratingNoiseSensitivity2006}. The definition's namesake $\epsilon$ is a positive number that indicates the acceptable degree of privacy loss allowed by the algorithm for a given individual whose data appears in the dataset. The smaller the value of $\epsilon$ the more similar the results must be to satisfy the required level of privacy. I directs readers to Chapter 2 of Dwork's and Roth's book on differential privacy \cite{dworkAlgorithmicFoundationsDifferential2014a} for the mathematically rigorous definition of differential privacy. Additionally, there are two key aspects of differential privacy: the requirement of the randomized algorithm, and the purpose of the privacy budget $\epsilon$. 

As the ability to maintain privacy in database queries requires plausible deniability of any outcome, if an algorithm is deterministic, it is impossible to ensure that the probabilities of the two results can be indistinguishable while still being useful \cite{dworkAlgorithmicFoundationsDifferential2014a}. To prevent this, a differentially private algorithm must add a random noise effect to perturb the resulting data. While this does sacrifice accuracy for privacy, this accuracy-privacy trade-off cannot be avoided when querying datasets for potentially sensitive information \cite{dworkAlgorithmicFoundationsDifferential2014a}.

Repeated iterations of these algorithms on a dataset begin to tear down the efficacy of the dataset's privacy. Despite the noisy nature of the algorithms, a statistical analysis can reveal the true data for a user contained within a dataset after sufficient repetition. The privacy budget $\epsilon$ for a given user serves as an upper bound to the allowed privacy leakage due to repeated dataset queries. The execution of a query on a dataset that contains a given user's data requires an amount of that user's privacy budget determined by the maximum distance between the result of the query on the true dataset and the result of the query on the true dataset minus that user's entry. If the required amount of budget exceeds the allowed budget, the query cannot be run and the algorithm returns with a failure; otherwise, the query result is returned and the budget used for each user's data contained within that query is subtracted from the total allowed budget for each respective user. Since each successful query reduces the allowed privacy budget for all users whose data is present in the dataset, there is a strictly finite number of queries that can be executed over the life of the dataset.

The authors of \cite{zhangDifferentialPrivacyCollaborative2021} seek to use differential privacy in a collaborative neural network environment. While it is common to train the models locally and provide only gradient updates across the network, as stated in Section \ref{ec-overview}, privacy leakage can still occur by using these gradient updates to reconstruct the original training data. By adding Gaussian noise to only the first layer of the neural network, the authors were able to achieve a model accuracy of 93\% while still satisfying $\epsilon < 1.4$ differential privacy requirements by reducing the overall privacy budget costs per epoch when compared to adding noise to every layer.

Local differential privacy (LDP) is a variation of differential privacy in which the user is responsible for perturbing their own entry before it is entered into the dataset. \cite{biPrivacypreservingMechanismBased2020} proposes that an LDP-based location algorithm can provide a greater quality of data without sacrificing privacy by decomposing the tracked area into a Voronoi diagram \cite{kangVoronoiDiagram2008} containing a set of polygonal zones. When asked to provide their location, a user device generates an array containing both it's real location and a set of fake locations within the polygon in which they currently reside. The device then selects a location to provide from the array by following the locally differentially private random response mechanism \cite{warnerRandomizedResponseSurvey1965a}. Initial experimental results show that the proposed LDP mechanism provides a higher quality of service than adding either Laplace or Gaussian noise to the data while ensuring the edge nodes cannot determine absolute location data for a user. Wang et. al also investigate the applications of LDP in edge computing privacy as it relates to task offloading in connected vehicles (CVs) \cite{wangJointOptimizationTask2022}. They find that disturbing actual CV speed and velocity by randomly selecting values within an interval calculated using an MWEM algorithm \cite{hardtSimplePracticalAlgorithm2012} not only provides a significant reduction in edge node task unloading processing delay when compared to the standard random response mechanism, but also retains its efficiency when tightening the privacy budget constraints.

A method for task allocation without LDP is discussed in \cite{wangAccuratePrivacyPreservingTask2022}. In this method, end devices provide real location data to edge nodes using pseudonyms. As the devices regenerate these pseudonym after each batch of task offloading, the edge nodes remain unable to link historical task data to a particular device. The location data from the edge nodes, in addition to the task data from the task allocation server, is then disturbed using a differentially private JL transformation \cite{yangDifferentialPrivatePOI2018} before being sent to an honest-but-curious third party. The third party may then allocate the tasks that, in turn, can then be provided to the edge nodes and performed by the end devices. When compared with standard methods of task encryption, the authors' results show both a large increase in task allocation precision and a large reduction in overall computational overhead while preserving end device location privacy.

\input{tables/data-uses-two}

\subsection{Secret Sharing}
Secret sharing schemes (also known as secret splitting schemes) split a secret, such as an decryption key, into $n$ shares which allow for the reconstruction of the original secret given a certain number of shares $k$. These shares are then distributed among authorized devices and users. Though secret sharing schemes which require all shares to reconstruct the secret (i.e. $k = n$) exist, many schemes only require $k$ shares to reconstruct the secret where $1 \leq k < n$ \cite{beimelSecretSharingSchemesSurvey2011}. The second category of schemes are known as $(k, n)$-threshold schemes; however, the $(k, n)$ notation may be used by schemes where $k = n$ to indicate that all shares are required to reconstruct the secret \cite{beimelSecretSharingSchemesSurvey2011} \cite{r.divyaSecretSharingSchemes2018}.

A critical property that must be present for a secret sharing scheme to be secure is the property that the shares must be generated in such a way that anyone possessing less than $k$ shares gains no more information about the secret than someone possessing no shares. Failure to ensure this property weakens the security of the system and risks exposing the secret to less than $k$ colluding shareholders with each additional colluding shareholder further reducing the effort required to reconstruct the original secret \cite{beimelSecretSharingSchemesSurvey2011}.

Model parameter sharing in collaborative learning is a primary use of secret sharing in current literature. The authors of \cite{hsuPrivacyPreservingFederatedLearning2020} use additive secret sharing to facilitate privacy in training malware detection models. In their implementation, after training a local model on their local data, the clients agree on a large prime then generate a random value with the same dimensions as the model parameters for each other client in the system. The clients then subtract each random value from their model's trained parameters to produce another value and distribute both the calculated and random values as shares. The clients can then send their shares to the edge node which calculates an aggregate, global model parameter using the additive properties of the shares. In this method, the clients' parameter updates are secure as only the clients are aware of the value of the prime and, therefore, only they reconstruct the true model parameters from the shares. In \cite{senguptaSPRITEScalablePrivacyPreserving2022}, the authors use a homomorphically additive secret sharing scheme which allows the end devices to split a gradient calculated from it's local model into shares distributed among the other end users and compute the summation of all of its received shares without altering the end result. The end devices then provide their summations to the edge node which can reconstruct the aggregated gradients without knowledge of a particular end device's gradient.

Ma et. al examine both model creation and prediction in \cite{maLightweightPrivacypreservingMedical2020}. After privately and securely generating an XGBoost powered decision tree \cite{nvidiaWhatXGBoost} on the system's edge nodes, end devices can perform private medical diagnosis by encrypting their symptom vector with a public key and submitting this vector, along with an encrypted random number, to the edge node. The edge node is then able to perform the decision tree classification by using a share of the private key to perform private comparison without being able to decrypt symptom data. Once a final diagnosis is made, the edge node passes the encrypted diagnosis multiplied by the encrypted random number back to the end device; the encrypted random number ensures only the proper end device is able to decrypt the final diagnosis so no diagnosis data is leaked to eavesdropping parties. In the energy sector, \cite{wangLiPSGLightweightPrivacyPreserving2020} performs both private training and recommendation through Q-learning \cite{watkinsQlearning1992}. By splitting Q-values of energy grid states and actions into separate shares of a Q-table, both private control center action recommendations and private Q-value updates can occur through additive share manipulations like those listed above. All the above papers saw virtually identical accuracy when compared with equivalent, non-private approaches.

Share-based operations are also used in image processing. The researchers in \cite{huangLightweightPrivacyPreservingCNN2021} privately extract image features by splitting the images into two shares before processing through a neural network. The layers requiring comparisons between image data can then use the shares to perform most-significant bit calculations which are used, in turn, to perform the required overall comparisons. Likewise, in \cite{yanPrivacypreservingContentbasedImage2022}, Yan .et al proposes an image storage and retrieval scheme in which a data owner can encrypt images symmetrically and upload them along with share-split feature information to two separate edge servers. When a user wishes to retrieve images, they request the decryption key and trapdoor function from the data owner, split the trapdoor function into shares, and provide the share-split trapdoor to the edge nodes. The nodes use a modified version of Du-Atallah protocol \cite{duProtocolsSecureRemote2001} to find encrypted images with features within a set distance of the query based on the trapdoor and return those images to the user, who possesses the symmetric key for decryption. As with the machine learning applications, these algorithms are no less accurate than their non-private counterparts and additionally remain negligibly less time efficient.

The authors of \cite{guSelfVerifiableAttributeBasedKeyword2022} provide a more abstracted scheme to process arbitrary data contents through distributing computed ciphertext shares through the entire network of an arbitrary number of edge nodes. Like the image retrieval algorithm, the data is broken into shares according to a public key and distributed across edge nodes. When an end device provides a trapdoor query to its edge node, the edge node coordinates searching all other edge nodes’ databases for ciphertext which matches the trapdoor. These ciphertexts are gathered on the directly connected edge node, aggregated, then sent to the requesting end device; the end device is then capable of recovering the original data by using a secret key. Unlike in \cite{yanPrivacypreservingContentbasedImage2022}, however, a TTP is required for key generation. Both algorithms provide improved storage and retrieval time efficiency when compared to other encrypted data retrieval algorithms as the heavy computations are delegated to the edge nodes, which are more powerful than any given user device.

Researchers are also exploring other applications of secret sharing in honest-but-curious edge computing such as Schlegel et. al’s \cite{schlegelPrivacyPreservingCodedMobile2022} mechanism of offloading linear computations from end devices with minimal computation power to more powerful edge nodes using split multiplication vectors and matrices and Zhou et. al’s algorithm \cite{zhaoDynamicPrivacyPreservingReputation2019} for crowdsensing observation blinding and aggregate reputation updating utilizing additive secret sharing.

\input{tables/crowdsense-source}

\subsection{Other Notable Techniques}
In addition to the main, recurring techniques above, several researchers have proposed algorithms based on less widely used techniques. The following subsections survey some of these techniques with the intent to both highlight promising opportunities for future research in less widely known areas and to provide background on possible research pitfalls.

\subsubsection{Slot Reservation}
Slot reservation for anonymous communication was originally proposed by Yao, Yang, and Xiong in \cite{yaoAnonymityBasedPrivacyPreservingData2015} to allow end devices to anonymously transmit observed data to a central server without key-based encryption. Under their original algorithm, each pair of end devices on the network share a pre-shared secret seed. Next, each end device would create a reservation message consisting of a chosen pseudonym and the length of the data they wished to transmit, encrypt this message using the public key of each end device in their system (including their own) in reverse order of a defined ordering (such that the last node was the innermost encryption), and send their messages to the first device in the ordering. Upon receiving all reservation messages, the first device would permutate the messages, strip off the first layer of encryption, and transmit the new permuted vector of messages to the second node. This process would continue until the final device possessed a final permutation of reservation messages with each layer of encryption stripped. Since the only information the last device holds is the permuted messages and its own pseudonym, it cannot determine where any other device’s messages lie in the vector. The device would then send this final vector to a central server which would broadcast it to all end devices.

From there, each end device could anonymously transmit data by taking advantage of the reversible properties of the exclusive or (XOR) operation. To accomplish this, an end device would generate pseudorandom concatenations of bit streams for each slot and each other device based on the pre-shared seeds, the slot owner’s pseudonym, and a predictable nonce. It would then perform an XOR on these concatenations per slot, XOR the observed data it wished to transmit with the final bitstream in its own pseudonym-keyed slot, pack each slot’s bit stream into a data array, and transmit the array to a central server. 

Once the central server received a data array from each end device, it would perform an XOR on each slot’s data across all arrays. Due to the reversibility of the XOR, the generated concatenations would cancel out, leaving only the observed data linked to a pseudonym which neither the end devices nor the server can use to identify the originator of the data. An example of this process involving three end devices who have reserved slots two, three, and one respectively is provided in Figure \ref{figure:slot-reservation-packing}.

\input{figures/slot-reservation-figure}

Though the slot reservation procedure itself was not created with edge computing in mind, the authors of \cite{liuPrivacypreservingRawData2019} extend this implementation to include the benefits of edge computing. In addition to adding an edge node as an intermediary between the end devices and the cloud server, their proposed algorithm does not require full communication between all end devices in the network, thereby drastically reducing the communication overhead required by the protocol. Additionally, they incorporate a signature verification mechanism to ensure the communication between the end devices, the edge nodes, and the cloud server is valid.

\subsubsection{Stochastic Location Cloaking}
The authors of \cite{tianStochasticlocationPrivacyProtection2020} propose a method for evaluating cloaked user locations against a configurable privacy threshold based on k-anonymity \cite{samaratiGeneralizingDataProvide1998}. The threshold involves the number of end devices a cloaked location area should contain ($k$), the minimum acceptable probability that the cloaked area contains that number of end devices ($P_k$), and a bounding on the size of the area to ensure the provided cloaked location is still useful for data processing. When an end device wishes to share its location, it generates a cloaked location based on the historical location data. A stochastic algorithm is then followed using current location data to ensure that the number of devices presently within the cloaked location is at least $k$ with probability $P_k$. If $P_k$ holds, the cloaked location is provided, otherwise the location data cannot be shared until a new cloaked location has been generated and validated. 

\subsubsection{End Device Coordination}
Jiang et. al \cite{jiangEfficientPrivacypreservingDistributed2021} proposes a gradient updating algorithm for federated learning which does not rely on complicated cryptographic mechanisms. By having each end device generate a random value and transmit that value to other end devices within its network, the gradient update of a single node can be blinded before being transmitted to an edge node by adding the sum of its sent values and subtracting the sum of its received values. The edge nodes then compute the sum of these blinded gradients, and further blind these aggregations in the same manner. Finally, the cloud server generates the new model gradient as the average of the summation of the received aggregations. Since, under this method, each generated value will be involved in an equal quantity of addition and subtraction operations, each one will cancel out during the summation operations leaving behind the true gradient average.

The downfall of this method that drives researchers to other techniques, such as homomorphic encryption, is the communication inefficiency. Each end device within a network must perform two-way communication between themselves which is prone to communication faults and can quickly become expensive, especially when considering the privacy protections which must be implemented over that network.

\subsubsection{Anonymous Communication through Short-Term Public Keys}
In their paper, Ernest and Shiguang \cite{ernestPrivacyEnhancementScheme2020} address the increasing reliance on communication through insecure channels that edge computing will likely bring. As more edge nodes are added to networks, the cost to secure all communication between devices owned by multiple entities will grow, and thus it is likely that more insecure channels will be introduced. This will necessitate a means for ensuring private communication across increasingly common insecure channels. Focusing on the blockchain, Ernest and Shiguang propose a method for per-transaction public key generation through elliptic curve cryptography \cite{gajbhiyeSurveyReportElliptic2011} given a chosen identity and private key. Their method allows an end device to encrypt its identity within a digital signature based on a quickly expiring public key. This allows for mutual identification and authentication between two end devices without leaking the identity information to the edge nodes involved in communication. Furthermore, as the public keys expire quickly and no identification information is provided during the broadcast of public keys to the network, edge nodes are incapable of linking communication paths between end devices beyond, at most, a few transactions.

\subsubsection{Probability-Based Offloading}
A task unloading method proposed by Zhu et. al \cite{zhuPrivacyAwareOnlineTask2021} sacrifices a small degree of system performance in order to perform offloading directly without a TTP or complicated encryption. After the end device identifies valid edge nodes capable of processing a task, it calculates the estimated cost for offloading the task to each given edge node. Instead of selecting the best choice, it weights the cost by a probability distribution tuned by a user’s privacy requirement in order to ensure edge nodes are not able to infer device locations based on offloading frequency. In addition to the probability calculations, end devices issue fake ``dummy'' tasks. These ensure the edge nodes are unable to determine the actual task request frequency at the cost of slight system performance degradation.

%% file: tables/techniques.tex
\begin{table*}[t]
\centering
\caption{Techniques Used in Surveyed Works}
{\renewcommand{\arraystretch}{1.2}%
\begin{tabular}{|c|c|c|c|c|c|c|c|c|c|c|c|c|c|c|c|}
\cline{2-16}
\multicolumn{1}{c|}{} & \cite{zengMMDAMultidimensionalMultidirectional2020} & \cite{hsuPrivacypreservingDataSharing2020} & \cite{hsuReconfigurableSecurityEdgeComputingBased2018} & \cite{liPrivacyPreservingData2019} & \cite{maEdgeComputingAssisted2021} & \cite{zhangLPDAECLightweightPrivacyPreserving2018} & \cite{liLightweightFineGrainedPrivacyPreserving2021} & \cite{liVerifiablePrivacypreservingMachine2021} & \cite{liPrivacyPreservedFederatedLearning2021} & \cite{dingPrivacypreservingTaskAllocation2022} & \cite{maPrivacyPreservingReputationManagement2019} & \cite{zhangDifferentialPrivacyCollaborative2021} & \cite{biPrivacypreservingMechanismBased2020} & \cite{wangJointOptimizationTask2022} & \cite{wangAccuratePrivacyPreservingTask2022} \\
\hline
Trusted Third-Party & \checkmark & \checkmark & \checkmark & & \checkmark & \checkmark & \checkmark & & \checkmark & & \checkmark & & \checkmark & & \\
\hline
Homomorphic Encryption & & & & \checkmark & \checkmark & \checkmark & \checkmark & \checkmark & \checkmark & \checkmark & \checkmark & & & & \\
\hline
Differential Privacy & & & & & & & & & & & & \checkmark & \checkmark & \checkmark & \checkmark \\ 
\hline
Secret Sharing & & & & & & & & & & & & & & & \\ 
\hline
Slot Reservation & & & & & & & & & & & & & & & \\ 
\hline
Stochastic Cloaking & & & & & & & & & & & & & & & \\ 
\hline
Device Coordination & & & & & & & & & & & & & & & \\ 
\hline
Short-Term Keys & & & & & & & & & & & & & & & \\ 
\hline
Probability-Based Offloading & & & & & & & & & & & & & & & \\ 
\hline
\end{tabular}}
\label{table:techniques:1}
\end{table*}

%% file: tables/techniques-two.tex
\begin{table*}[t]
\centering
\caption{Techniques Used in Surveyed Works (continued)}
{\renewcommand{\arraystretch}{1.2}%
\begin{tabular}{|c|c|c|c|c|c|c|c|c|c|c|c|c|c|c|}
\cline{2-15}
\multicolumn{1}{c|}{} & \cite{hsuPrivacyPreservingFederatedLearning2020} & \cite{senguptaSPRITEScalablePrivacyPreserving2022} & \cite{maLightweightPrivacypreservingMedical2020} & \cite{wangLiPSGLightweightPrivacyPreserving2020} & \cite{huangLightweightPrivacyPreservingCNN2021} & \cite{yanPrivacypreservingContentbasedImage2022} & \cite{guSelfVerifiableAttributeBasedKeyword2022} & \cite{schlegelPrivacyPreservingCodedMobile2022} & \cite{zhaoDynamicPrivacyPreservingReputation2019} & \cite{liuPrivacypreservingRawData2019} & \cite{tianStochasticlocationPrivacyProtection2020} & \cite{jiangEfficientPrivacypreservingDistributed2021} & \cite{ernestPrivacyEnhancementScheme2020} & \cite{zhuPrivacyAwareOnlineTask2021} \\
\hline
Trusted Third-Party & \checkmark & \checkmark & \checkmark & \checkmark & \checkmark & \checkmark & \checkmark & & \checkmark & & & & & \\
\hline
Homomorphic Encryption & & & \checkmark & & & & & & & & & & & \\ 
\hline
Differential Privacy & & & & & & & & & & & & & & \\ 
\hline
Secret Sharing & \checkmark & \checkmark & \checkmark & \checkmark & \checkmark & \checkmark & \checkmark & \checkmark & \checkmark &  & & & & \\ 
\hline
Slot Reservation & & & & & & & & & & \checkmark & & & & \\ 
\hline
Stochastic Cloaking & & & & & & & & & & & \checkmark & & & \\ 
\hline
Device Coordination & & & & & & & & & & & & \checkmark & & \\ 
\hline
Short-Term Keys & & & & & & & & & & & & & \checkmark & \\ 
\hline
Probability-Based Offloading & & & & & & & & & & & & & & \checkmark \\ 
\hline
\end{tabular}}
\label{table:techniques:2}
\end{table*}

%% file: tables/data-uses.tex
\begin{table*}[t]
\centering
\caption{Data Processing Paradigms in Surveyed Works}
{\renewcommand{\arraystretch}{1.2}%
\begin{tabular}{|c|c|c|c|c|c|c|c|c|c|c|c|c|c|c|c|}
\cline{2-16}
\multicolumn{1}{c|}{} & \cite{zengMMDAMultidimensionalMultidirectional2020} & \cite{hsuPrivacypreservingDataSharing2020} & \cite{hsuReconfigurableSecurityEdgeComputingBased2018} & \cite{liPrivacyPreservingData2019} & \cite{maEdgeComputingAssisted2021} & \cite{zhangLPDAECLightweightPrivacyPreserving2018} & \cite{liLightweightFineGrainedPrivacyPreserving2021} & \cite{liVerifiablePrivacypreservingMachine2021} & \cite{liPrivacyPreservedFederatedLearning2021} & \cite{dingPrivacypreservingTaskAllocation2022} & \cite{maPrivacyPreservingReputationManagement2019} & \cite{zhangDifferentialPrivacyCollaborative2021} & \cite{biPrivacypreservingMechanismBased2020} & \cite{wangJointOptimizationTask2022} & \cite{wangAccuratePrivacyPreservingTask2022} \\
\hline
Transmission & & \checkmark & \checkmark & & & & & & & & & & \checkmark & & \\
\hline
Aggregation & \checkmark & & & \checkmark & \checkmark & \checkmark & \checkmark & & & & & & & & \\
\hline
Federated Learning & & & & & & & & \checkmark & \checkmark & & & \checkmark & & & \\ 
\hline
Offloading & & & & & & & & & & \checkmark & \checkmark & & & \checkmark & \checkmark \\ 
\hline
\end{tabular}}
\label{table:data-uses:1}
\end{table*}

%% file: tables/data-uses-two.tex
\begin{table*}[t]
\centering
\caption{Data Processing Paradigms in Surveyed Works (continued)}
{\renewcommand{\arraystretch}{1.2}%
\begin{tabular}{|c|c|c|c|c|c|c|c|c|c|c|c|c|c|c|}
\cline{2-15}
\multicolumn{1}{c|}{} & \cite{hsuPrivacyPreservingFederatedLearning2020} & \cite{senguptaSPRITEScalablePrivacyPreserving2022} & \cite{maLightweightPrivacypreservingMedical2020} & \cite{wangLiPSGLightweightPrivacyPreserving2020} & \cite{huangLightweightPrivacyPreservingCNN2021} & \cite{yanPrivacypreservingContentbasedImage2022} & \cite{guSelfVerifiableAttributeBasedKeyword2022} & \cite{schlegelPrivacyPreservingCodedMobile2022} & \cite{zhaoDynamicPrivacyPreservingReputation2019} & \cite{liuPrivacypreservingRawData2019} & \cite{tianStochasticlocationPrivacyProtection2020} & \cite{jiangEfficientPrivacypreservingDistributed2021} & \cite{ernestPrivacyEnhancementScheme2020} & \cite{zhuPrivacyAwareOnlineTask2021} \\
\hline
Transmission & & & & & & \checkmark & \checkmark & \checkmark & & \checkmark & \checkmark & & \checkmark & \\
\hline
Aggregation & & & & & & & & & \checkmark & & & & & \\ 
\hline
Federated Learning & \checkmark & \checkmark & \checkmark & \checkmark & \checkmark & & & & & & & \checkmark & & \\ 
\hline
Offloading & & & & & & & & & & & & & & \checkmark \\ 
\hline
\end{tabular}}
\label{table:data-uses:2}
\end{table*}

%% file: tables/crowdsense-source.tex
\begin{table*}[t]
\centering
\caption{Surveyed Works intended for Crowdsensing}
{\renewcommand{\arraystretch}{1.2}%
\begin{tabular}{|c|c|}
\hline
Papers \\ 
\hline
\cite{dingPrivacypreservingTaskAllocation2022}, \cite{maPrivacyPreservingReputationManagement2019}, \cite{wangAccuratePrivacyPreservingTask2022}, \cite{zhaoDynamicPrivacyPreservingReputation2019}\\
\hline
\end{tabular}}
\label{table:crowdsense-source}
\end{table*}

%% file: figures/slot-reservation-figure.tex
\begin{figure}[ht]
    \centering
    \includegraphics[width=0.3\textwidth]{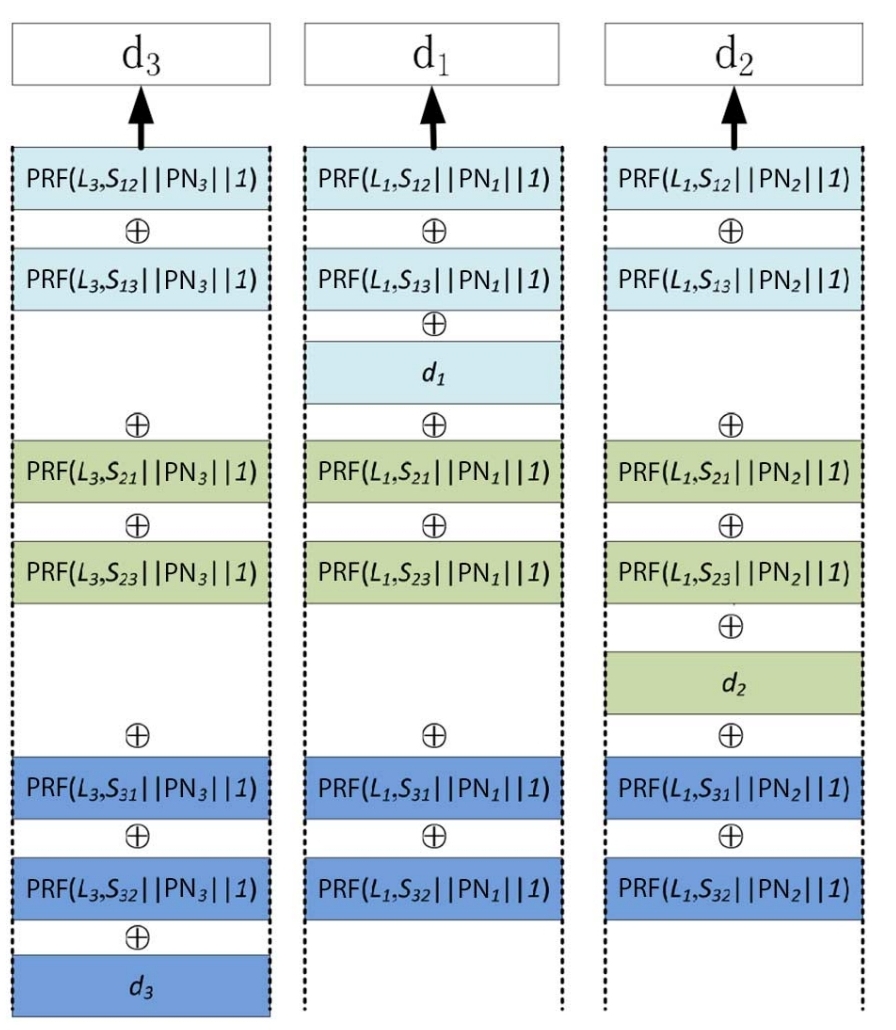}
    \caption{An example of slot reservation where PRF is a pseudorandom  function, $L_i$ is the length of the data in slot $i$, $S_{ij}$ is a shared secret between devices $i$ and $j$, PN$_i$ is the pseudonym of device $i$, and $d_i$ is device $i$'s observed data}
    \source{\textit{Anonymity-Based Privacy-Preserving Data Reporting for Participatory Sensing}, by Y. Yao, L. T. Yang, and N. N. Xiong, 2015, \textit{IEEE Internet of
Things Journal}}
    \label{figure:slot-reservation-packing}
\end{figure}

%% file: conclusion.tex
\section{Conclusion} \label{conclusion}

As more network architectures become defined by edge computing due to its potential for reduced requirements on bandwidth, decreased latency, increased scalability, and increased processing power, it becomes critical for those at the forefront of edge computing design and implementation to incorporate the protection of users' privacy as a foundational element in their algorithms. Moving forward, research should focus on eliminating the need for TTPs while still respecting the limited resources present on IoT devices as TTPs can inject a single point of failure into otherwise robust systems. Communication overheads must also be minimized since the bandwidth required for excessive back-and-forth communication may not be obtainable for all network owners as their edge networks grow. Finally, researchers and algorithm designers must develop unobtrusive, user-transparent implementations in order to effectively address even the most novice user's concerns, earn their trust, and protect their data. 

In this paper, I start by examining in which ways privacy differs from security. I then introduce edge computing as a network architecture paradigm and provide examples of application domains and data processing paradigms in which edge computing is actively used. Finally, I conducted a survey of current literature on privacy protection in honest-but-curious edge computing and categorized the papers by the overarching techniques used within their proposed algorithms as well as the authors' data processing goals. I conducted this survey to emphasize both the importance and practically of data privacy protection within edge computing in the hope that future research will expand upon existing work to ensure a future in which users can enjoy the benefits provided by state-of-the-art edge computing architectures while remaining confident that the privacy of their data is safeguarded.